\documentclass{moriond}

\newcommand{\Photo}{}

\usepackage{amsmath}
\usepackage{xspace}
\usepackage{ifthen}
\usepackage{siunitx}
\usepackage{cleveref}
\usepackage{bm}
\usepackage{lineno}
\usepackage{multirow}

\def\RKresult{\mbox{\ensuremath{0.846\,^{+\,0.042}_{-\,0.039}\,^{+\,0.013}_{-\,0.012}}}}

\def\RKsiggers{\mbox{\ensuremath{\SI{3.1}{\sigma}}}\xspace}

\newcommand{\JPsi}{{\ensuremath{J/\psi}}\xspace}

\newcommand{\PsiS}{{\ensuremath{\psi(2S)}}\xspace}

\newcommand{\Jpsi}{{\ensuremath{J/\psi}}\xspace}

\newcommand{\BKll}{\mbox{\ensuremath{B^+ \rightarrow K^+ \ell^+ \ell^-}}\xspace}

\newcommand{\BKmumu}{\mbox{\ensuremath{B^+ \rightarrow K^+ \mu^+ \mu^-}}\xspace}
\newcommand{\BKee}{\mbox{\ensuremath{B^+ \rightarrow K^+ e^+ e^-}}\xspace}

\newcommand{\BKstee}{\mbox{\ensuremath{B^0 \rightarrow \Kstarz(K^+\pi^-) e^+ e^-}}\xspace}

\newcommand{\BHee}{\mbox{\ensuremath{H_b \rightarrow H e^+ e^-}}\xspace}
\newcommand{\BHmumu}{\mbox{\ensuremath{H_b \rightarrow H \mu^+ \mu^-}}\xspace}

\newcommand{\BKJpsill}{\mbox{\ensuremath{B^+ \rightarrow K^+ J/\psi(\ell^+ \ell^-)}}\xspace}

\newcommand{\Kll}{\ensuremath{K^+ \ell^+ \ell^-}\xspace}

\newcommand{\Kmumu}{\ensuremath{K^+ \mu^+ \mu^-}\xspace}

\newcommand{\Kee}{\ensuremath{K^+ e^+ e^-}\xspace}

\newcommand{\KJpsiee}{\ensuremath{K^+ J/\psi(e^+ e^-)}\xspace}
\newcommand{\KJpsimumu}{\ensuremath{K^+ J/\psi(\mu^+ \mu^-)}\xspace}
\newcommand{\KJpsill}{\ensuremath{K^+ J/\psi(\ell^+ \ell^-)}\xspace}

\newcommand{\KpsiSee}{\ensuremath{K^+ \psi(2S)(e^+ e^-)}\xspace}
\newcommand{\KpsiSmumu}{\ensuremath{K^+ \psi(2S)(\mu^+ \mu^-)}\xspace}

\newcommand{\btosll}{\ensuremath{b \to s \ell^+ \ell^-}\xspace}

\newcommand{\leplep}{{\ensuremath{\ell^+ \ell^-}}\xspace}
\newcommand{\ptr}{{\ensuremath{p_{\rm{T}}}}\xspace}

\newcommand{\eff}{\ensuremath{\varepsilon}}
\newcommand{\RK}{{\ensuremath{R_K}}\xspace}
\newcommand{\Rk}{{\ensuremath{R_K}}\xspace}
\newcommand{\RpK}{{\ensuremath{R_{pK}}}\xspace}
\newcommand{\RKst}{{\ensuremath{R_{K^{*0}}}}\xspace}

\newcommand{\rJpsi}{{\ensuremath{r_{\JPsi}}}\xspace}

\newcommand{\RPsiS}{{\ensuremath{R_{\psi(2S)}}}\xspace}
\newcommand{\RpsiS}{{\ensuremath{R_{\psi(2S)}}}\xspace}

\newcommand{\qsqRange}{\mbox{\ensuremath{\qsq\in(\SI{1.1}{\gev^2},\;\SI{6.0}{\gev^2})}}\xspace}

\newcommand{\mkee}{\ensuremath{{m(\Kee)}}\xspace}

\newcommand{\viz}{\mbox{\itshape viz.}\xspace}

\newcommand{\cf}{\mbox{\itshape cf.}\xspace}

\def\PB      {\ensuremath{{B}}\xspace}                 
\def\PK      {\ensuremath{{K}}\xspace}                 
\def\B       {{\ensuremath{\PB}}\xspace}
\def\Bu      {{\ensuremath{\B^+}}\xspace}
\def\Bd      {{\ensuremath{\B^0}}\xspace}
\def\kaon    {{\ensuremath{\PK}}\xspace}

\def\Kp      {{\ensuremath{\kaon^+}}\xspace}
\def\Km      {{\ensuremath{\kaon^-}}\xspace}

\def\Kstarz  {{\ensuremath{\kaon^{*0}}}\xspace}
\def\qsq       {{\ensuremath{q^2}}\xspace}
\def\BF         {{\ensuremath{\mathcal{B}}}\xspace}
\def\BR         {\BF}

\newcommand{\aunit}[1]{\ensuremath{\text{\,#1}}}       
  
\newcommand{\tev}{\aunit{Te\kern -0.1em V}\xspace}
\newcommand{\gev}{\aunit{Ge\kern -0.1em V}\xspace}
\newcommand{\mev}{\aunit{Me\kern -0.1em V}\xspace}
\newcommand{\kev}{\aunit{ke\kern -0.1em V}\xspace}
\newcommand{\ev}{\aunit{e\kern -0.1em V}\xspace}
\def\fb   {\ensuremath{\aunit{fb}}\xspace}
\def\invfb   {\ensuremath{\fb^{-1}}\xspace}

\begin{document}

\vspace*{4cm}
\title{TEST OF LEPTON FLAVOUR UNIVERSALITY IN \boldmath{\btosll}\unboldmath~DECAYS}

\author{R. D. MOISE}

\address{Blackett Laboratory, Imperial College London\\South Kensington Campus, London SW7 2AZ, United Kingdom}

\maketitle\abstracts{
The Standard Model of particle physics predicts that charged leptons have the same electroweak interaction strength.
This symmetry, called lepton flavour universality, was found to hold in a wide range of particle decays.
One observable that is sensitive to lepton flavour universality is the ratio of branching fractions $\RK=\BR(\BKmumu)\,/\,\BR(\BKee)$. This quantity is measured using \SI{9}{\invfb} of proton-proton collision data recorded by the LHCb experiment at CERN's Large Hadron Collider.
For the dilepton invariant mass squared range \qsqRange, the result is $\RK=\RKresult$, where the first uncertainty is statistical and the second systematic. The measured value is in tension with the Standard Model prediction at the level of \SI{3.1}{\sigma}, thus providing evidence for the violation of lepton flavour universality in \BKll decays.
}

\section{Introduction}

The Standard Model (SM) of particle physics offers the most precise predictions on the properties and interactions of fundamental particles. Although it has withstood experimental scrutiny on numerous fronts, there are still some observations not accounted for in the SM. Examples of these are the observed matter-antimatter imbalance on the cosmological scale, and the effects attributed to dark matter in the Universe. 
One way to extend the SM, such that it is able to explain these phenomena, is by searching for new particles and interactions, collectively referred to as ``new physics'' (NP).

\Cref{sec:btosll} introduces \btosll transitions, which are a particular set of processess susceptible to NP effects. This is followed by~\Cref{sec:anomalies}, which covers the current experimental status.~\Cref{sec:procedure} provides an overview of the experimental techniques employed by the measurement of \RK. Then,~\Cref{sec:result} presents the \RK result, and finally~\Cref{sec:conclusions} summarises the findings and their implications. Natural units, where $\hbar=c=1$, are used throughout, with the exception of the invariant mass plots. Charge conjugation is implied, unless stated otherwise.

\subsection{Rare beauty-quark decays}\label{sec:btosll}

\begin{figure}[tpb]
   \begin{center}
      \includegraphics[width=0.96\linewidth]{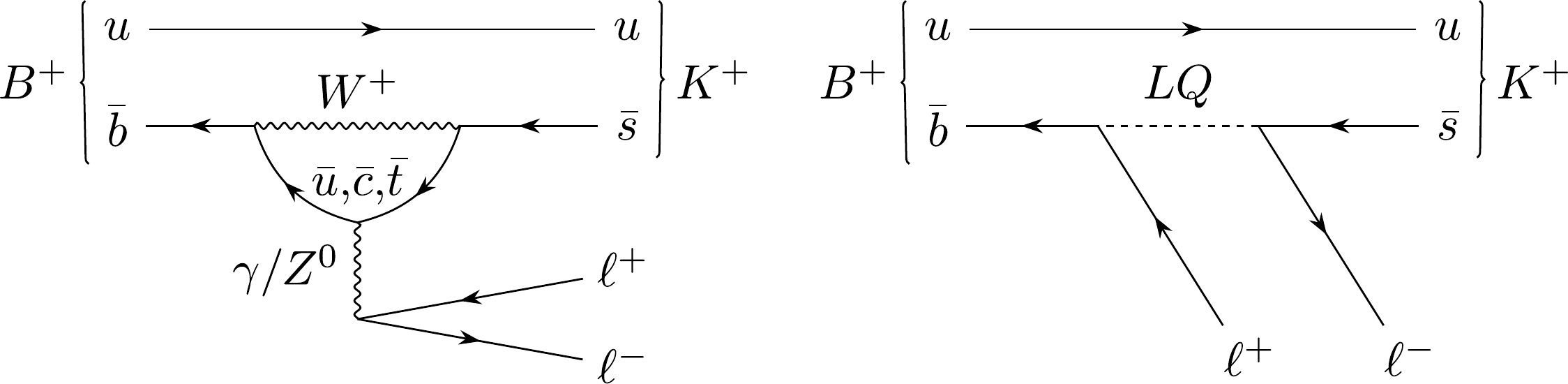}
   \end{center}
   \vspace*{-0.5cm}
   \caption{
   Loop-level Feynman diagram for the \BKll process (left). Because \btosll transitions are forbidden at tree level in the SM, this diagram is one of the leading contribution to this process in the SM. Possible NP, such as a leptoquark ($LQ$), could however allow this process to proceed at tree level (right).    
   }\label{fig:feyn}
\end{figure}

The SM is able to offer clear predictions of certain decay rates. Among these are processes in which a \Bu decays into a \Kp with the emission of a lepton-antilepton pair, \leplep. 
Because these processes involve the change of a $b$-quark into an $s$-quark, they are collectively referred to as \btosll transitions. In the SM, they are forbidden at tree level~\cite{PhysRevD.2.1285}, and therefore have to proceed via loop processes. This is illustrated by the left-hand side Feynman diagram in~\Cref{fig:feyn}. As a result of this suppression, the branching fractions of \BKll decays are expected to be low, \viz of $\mathcal O(10^{-6})$~\cite{PDG2020}.

However, such suppression does not necessarily apply to NP, as illustrated by the right-hand side Feynman diagram in~\Cref{fig:feyn}. Some models predict particles that could allow the \btosll transition to proceed at tree level. Such processes would therefore benefit from an enhancement with respect to their SM counterparts. This would result in a departure of the relative decay rate of such a process, also known as its branching fraction, away from its SM prediction. Therefore, precision measurements of \btosll transitions could be susceptible to such deviations, and are therefore promising areas of study in the search for NP. 

\subsection{Flavour anomalies}\label{sec:anomalies}

In the last decade, a pattern has emerged between experimental results on \btosll transitions~\cite{Altmannshofer:2021qrr}. These are collectively referred to as ``flavour anomalies''~\footnote{There exist flavour anomalies which are not examples of \btosll transitions. They are outside the scope of these proceedings, which focus on \btosll results.}, and they manifest themselves as tensions between measured observables related to beauty-quark decays, and their SM predictions. Such tensions are typically up to $2.9$~standard deviations. This does not conclusively prove the presence of NP, therefore futher studies are needed. The situation is complicated by the fact that predicting certain flavour observables in the SM is difficult, due to QCD effects which cannot be quantified using perturbation theory. 

However, leptons are not affected at leading-order by the strong interaction, and therefore QCD effects are expected to be indentical in, for example, \BKmumu and \BKee decays. This means that the following class of observables are theoretically clean:

\begin{equation}
\label{eq:rh}
R_H \equiv  \dfrac{\displaystyle\int_{q^2_\mathrm{min}}^{q^2_{\rm max}} \dfrac{\mathrm{d}\BF(\BHmumu)}{\mathrm{d}\qsq} \mathrm{d}\qsq}{\displaystyle\int_{q^2_{\rm min}}^{q^2_{\rm max}} \dfrac{\mathrm{d}\BF(\BHee)}{\mathrm{d}\qsq} \mathrm{d}\qsq} ~.
\end{equation}

\noindent Here, $R_H$ is a ratio of differential branching fractions, integrated over the dilepton invariant mass squared range $\qsq\in[q^2_\mathrm{min},\,q^2_{\rm max}]$. The initial-state particle, $H_b$, can be any hadron with a valence \mbox{$b$-quark}; examples include \Bu, \Bd, and $\Lambda_b^0$. The $H$ in the final state can be either a particle, such as a \Kp, or a system of particles like $pK^-$.

In the SM, such branching fraction ratios can be predicted with $\mathcal O(1\%)$ precision~\cite{GinoRKPrediction1Percent,Isidori:2020acz}.
This is thanks to an accidental symmetry of the SM, known as lepton flavour universality (LFU). It predicts that the three charged lepton families couple with the same interaction strength to vector bosons. This means that the ratios $R_H$ are expected to be close to unity, with small deviations induced by effects such as QED corrections, and negligible phase space differences arising from the different electron and muon masses.

The particular cases of $(H_b,\,H)=(\Bu,\,\Kp),\,(\Bd,\,\Kstarz)$, and $(\Lambda_b^0,\,p\Km)$ are known as \RK, \RKst, and \RpK, respectively. The latter has been measured~\cite{Aaij:2019bzx} and found to be compatible with unity at the level of \SI{1}{\sigma}. However, the most precise measurements of the other two ratios~\cite{LHCB-PAPER-2019-009,Aaij:2017vbb} exhibit tensions with the SM above \SI{2}{\sigma}. In particular, the LHCb collaboration~\cite{Alves:2008zz,LHCb-DP-2014-002} has measured \RK in the past, using \SI{5}{\invfb} of proton-proton collision data, collected up to and including the year 2016. This led to a result \SI{2.5}{\sigma} away from the SM prediction. These proceedings present an updated measurement of \RK from LHCb~\cite{Aaij:2021vac}, with the addition of data collected during the years 2017 and 2018. This corresponds to an approximate doubling of the data set with respect to the previous measurement.

\section{Procedure}\label{sec:procedure}

\subsection{The double-ratio strategy}\label{sec:doubleRatio}

One of the main challenges of conducting a measurement such as \Rk is the different behaviour of muons and electrons as they pass through matter. At LHCb, most muons are expected to traverse the detector nearly unimpeded, before being stopped by the muon stations. However, electrons may lose significant amounts of energy to bremsstrahlung radiation. This means that they have poorer mass resolution \cf muons, and that electron tracks are not resconstructed with the same efficiency as muon tracks. The event selection also has to be adapted, in order to take into account the different experimental signatures of the two lepton flavours. This makes \RK susceptible to systematic effects induced by differences in the muon and electron selections.

To suppress such systematic effects, \RK is measured with respect to well-known control channels. Resonant \BKJpsill decays are chosen for their large statistics and similar kinematics with respect to the \BKll modes, as well as for being known within $\mathcal O(10^{-4})$ to be compatible with LFU~\cite{PDG2020}. By taking the ratio between rare \BKll and control \BKJpsill branching fractions, written as ratios of yields and efficiencies, the experimental observable \RK is defined as:

\begin{equation}
\label{eq:RKyieldsEffs}
\Rk = \left(\frac{N(\Kmumu)}{\eff(\Kmumu)}\cdot\frac{\eff(\Kee)}{N(\Kee)}\right)\left/\middle(\frac{N(\KJpsimumu)}{\eff(\KJpsimumu)}\cdot\frac{\eff(\KJpsiee)}{N(\KJpsiee)}\right).
\end{equation}

\noindent In the above expression, $N(X)$ and $\eff(X)$ are, respectively, the yields and efficiencies of the decay of a \Bu into $X$. Thanks to the similar kinematics of the final-state particles from the rare and control decays, the selection is kept identical up to the cut on \qsq and on the invariant mass of the \Kll system. 
% These cuts are reported in~\Cref{tab:q2ranges}.
In particular, the rare-mode samples are selected in \qsqRange.
The rest of the selection ensures the good quality of the tracks and \Bu decay vertex, and rejects backgrounds using dedicated requirements such as mass vetoes and multivariate discriminators. The selection is essentially identical to the one employed by the previous \RK measurement at LHCb~\cite{LHCB-PAPER-2019-009}.

\subsection{Calculation of efficiencies}\label{sec:efficiencies}

Simulated events are used to estimate the efficiency of the selection. Some features are modelled imperfectly by the simulation, and therefore require calibration. This procedure is identical to the one employed in the previous \RK measurement~\cite{LHCB-PAPER-2019-009}, and it involves comparing the selection performance between control-mode data and simulation. Based on the differences found, weights are assigned to simulated events in order to align their performance to the one found in data. This procedure is applied to calibrate: the generated and reconstructed kinematics of the \Bu, the trigger and particle identification~\footnote{The calibration of particle identification performance follows the procedure described in Ref.~\cite{Aaij:2018vrk}.} performances, and the resolutions of \qsq and \mkee.

\subsection{Validation of the procedure}\label{sec:xchecks}

The experimental procedure is verified through several tests, each targeting specific aspects of the measurement strategy. Two such tests are presented here. The first one consists of measuring the single ratio \rJpsi~, defined as the denominator containing all \KJpsill terms in~\Cref{eq:RKyieldsEffs}. It represents a stringent test of the efficiencies, because it does not benefit from the cancellation of systematic effects --- due to electron and muon detection differences --- that is built into the double ratio \RK. Therefore, an accurate measurement of \rJpsi requires suitable understanding of both electron and muon absolute efficiencies. The average value~\footnote{The average takes into account the correlations between the different selections used in the analysis.} across the entire dataset is found to be \mbox{$\rJpsi=0.981\pm0.020$}, where the uncertainty contains both statistical and systematic uncertainties. This value is compatible within \SI{1}{\sigma} with the expectation from LFU. 

\begin{figure}[!htpb]
\begin{center}
\includegraphics[width=0.41\linewidth]{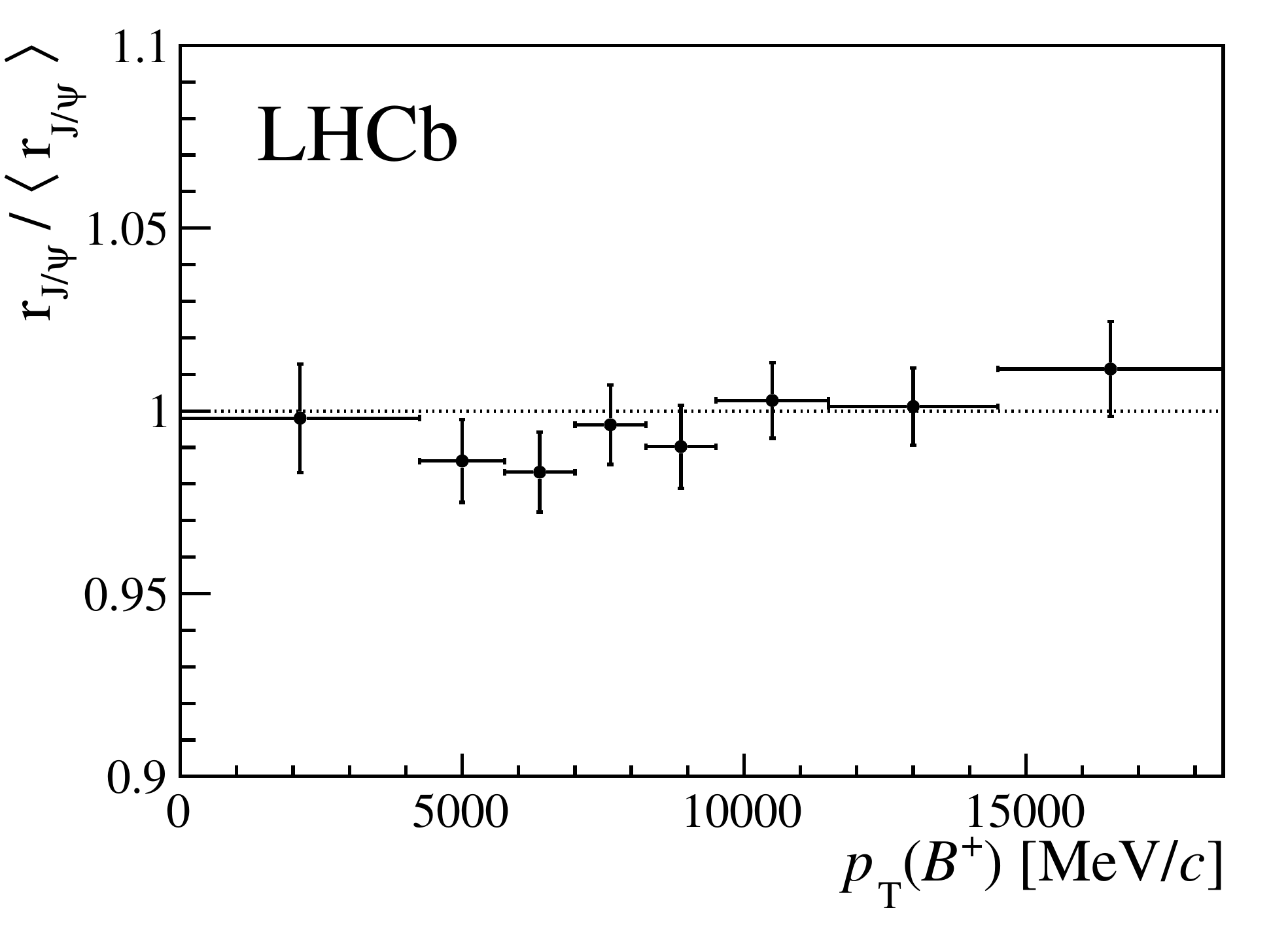}
\hspace{1em}
\includegraphics[width=0.41\linewidth]{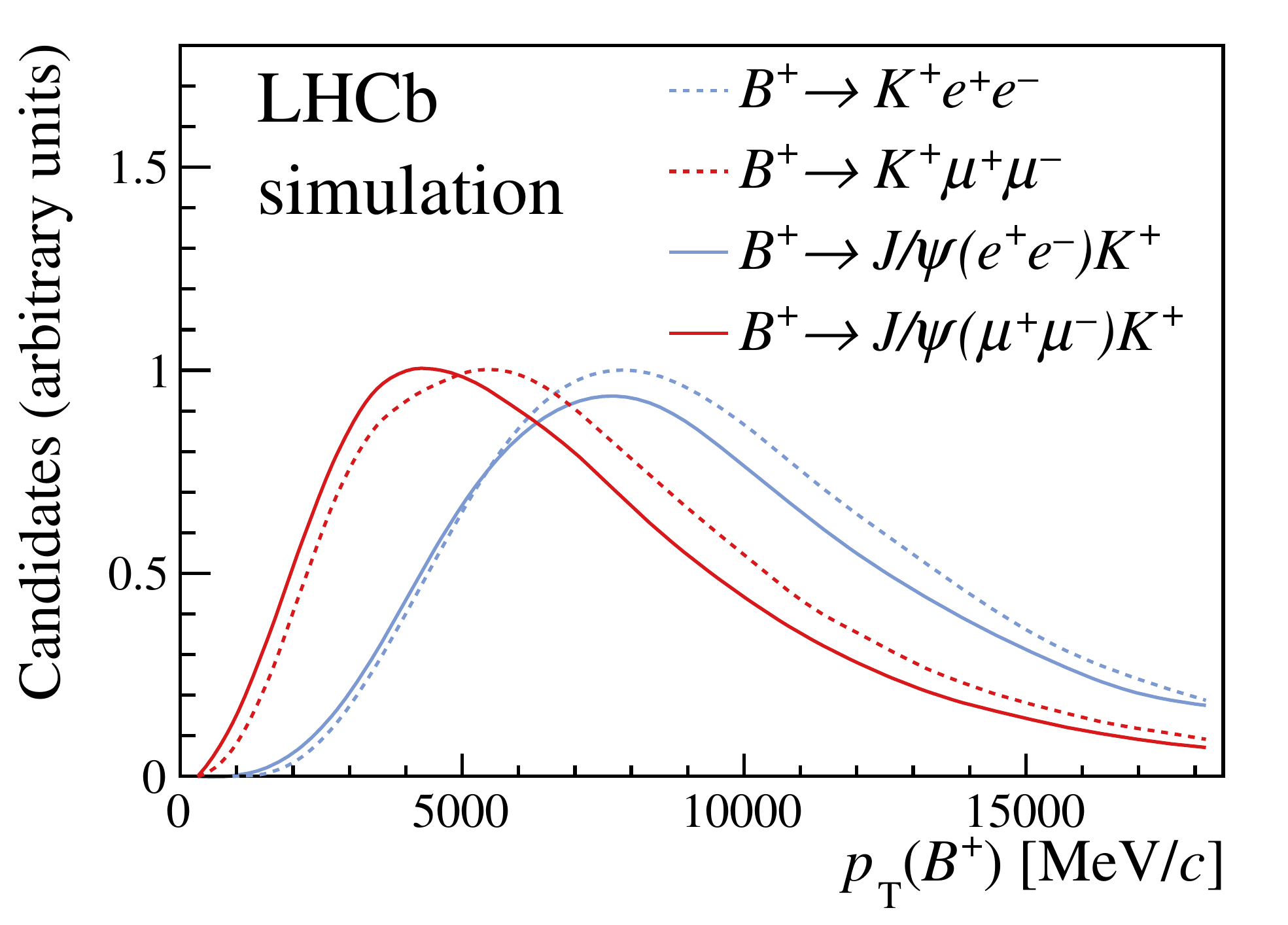}
\end{center}
\vspace*{-0.5cm}
\caption{
The single ratio \rJpsi, computed as a function of the transverse momentum of the \Bu (left). The values are normalised to their average. The distribution of this variable in rare- and control-mode simulation is shown on the right.
}\label{fig:1DrJpsi}
\end{figure}

To further test the efficiencies, \rJpsi is also calculated in bins of variables relevant to the detector response. 
In the case of ideal efficiency corrections, \rJpsi would show no dependence on any kinematic variable.
An example is shown in~\Cref{fig:1DrJpsi} on the left, where \rJpsi is found to be uniform across the transverse momentum of the \Bu, $\ptr(\Bu)$
. If any residual differences from unity are assumed to be caused by genuine mismodelling of the efficiencies, their effect on \RK can be estimated based on the differences in the $\ptr(\Bu)$ spectrum between the rare and control modes. As shown in~\Cref{fig:1DrJpsi} on the right, these differences are small, despite their separation in \qsq. The expected impact on \RK is found to be within the systematic uncertainty on \RK. This is found to be true for all considered kinematic variables. Dependencies on pairs of variables are also checked, and also show deviations from uniformity smaller than the applied systematic uncertainty.

The efficiencies are futher tested by calculating another double ratio, but centred on the \PsiS charmonium state:
\begin{align}
\label{eq:Rpsi2S}
\nonumber
\RPsiS = &\left(\frac{N(\KpsiSmumu)}{\eff(\KpsiSmumu)}\cdot\frac{\eff(\KpsiSee)}{N(\KpsiSee)}\middle)\right/\\
&\left(\frac{N(\KJpsimumu)}{\eff(\KJpsimumu)}\cdot\frac{\eff(\KJpsiee)}{N(\KJpsiee)}\right).
\end{align}

\noindent The \PsiS resonance is well separated in \qsq from the \Jpsi, which means \RpsiS probes the validity of the efficiency corrections across \qsq. In addition, \RpsiS is a double ratio, thus allowing the cancellation of systematic effects to be tested. The result is $\RpsiS=0.997\pm0.011$, where the uncertainty encompasses both statistical and systematic effects. This is compatible with the LFU prediction of unity, and demonstrates that systematic uncertainties on the double ratio due to electron-muon differences are expected to be below $1\%$.

\section{Result}\label{sec:result}

\begin{figure}[!htpb]
\begin{center}
\includegraphics[width=0.41\linewidth]{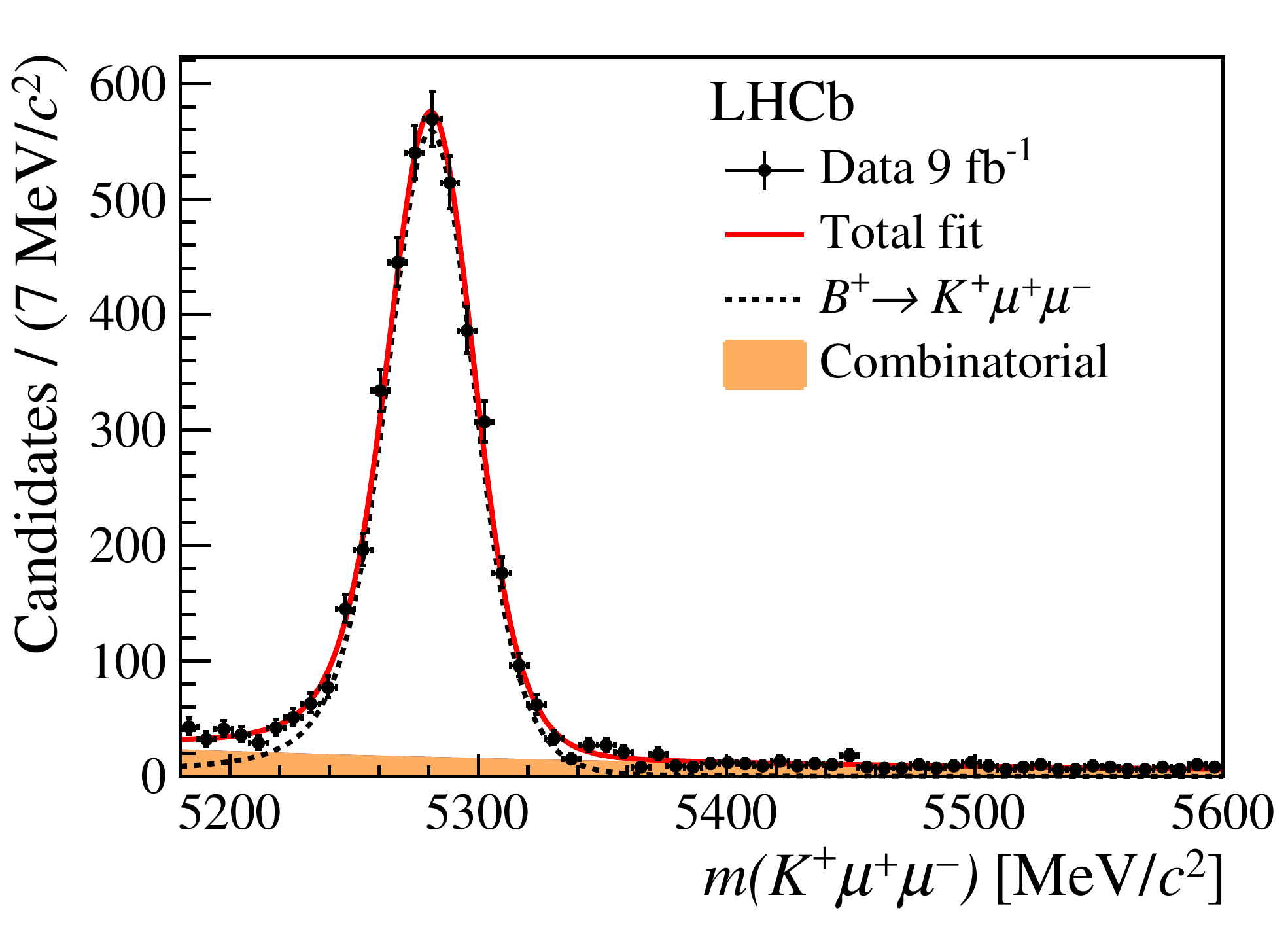}
\hspace{1em}
\includegraphics[width=0.41\linewidth]{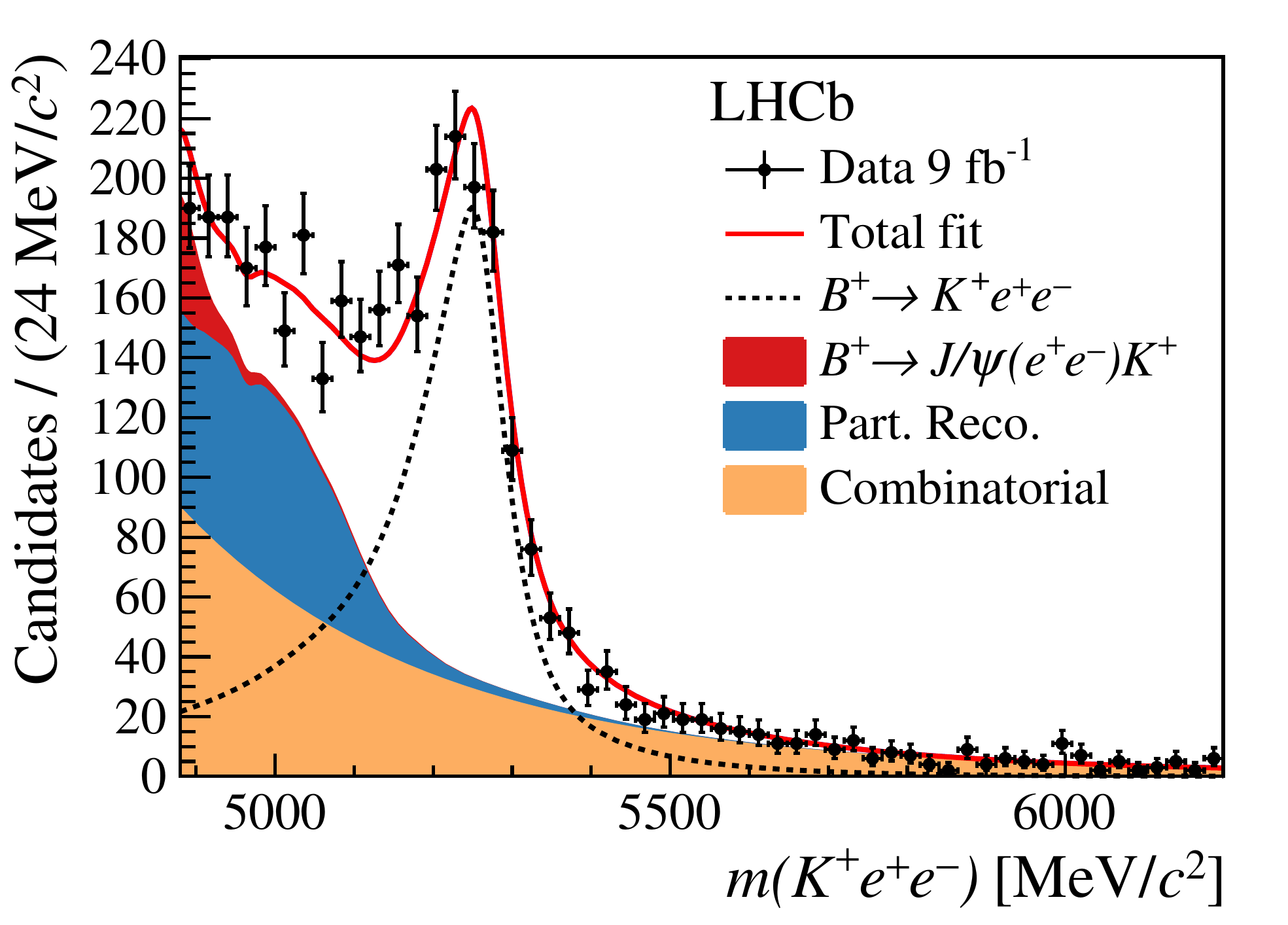}
\end{center}
\vspace*{-0.5cm}
\caption{
Projections of the fit to candidate \BKmumu (left) and \BKee (right) invariant-mass distributions. 
}\label{fig:RKfit}
\end{figure}

An unbinned extended maximum likelihood fit~\cite{BARLOW1990496} is performed simultaneously on \BKmumu and \BKee data. The fit projections are shown in~\Cref{fig:RKfit}. Calibrated simulation is used to determine the shapes of the fit components, alongside their relative normalisation. The correlations between the various selections used to obtain the data are taken into account. In both muon and electron data, the signal forms a peaking structure, which sits on top of a continuum consisting of so-called combinatorial background events. The latter correspond to candidates formed from random combinations of tracks in the event. Due to the poorer mass resolution, the low-mass tail of the electron signal overlaps with a contribution from the \Jpsi mode, and with a structure formed by partially-reconstructed background events. These consist of processes where kaons and a dilepton pair are formed, in addition to other particles that escape detection. An example of this is the \BKstee process, where the pion is not reconstructed.

The observable \RK is inferred as one of the fit's parameters of interest. The value from the fit is adjusted by correcting the central value based on the estimated bias of the fit, which is found to be small. The uncertainty is adjusted to account for the systematic uncertainty induced by the fit model.

\begin{figure}[tpb]
\begin{center}
\includegraphics[width=0.41\linewidth]{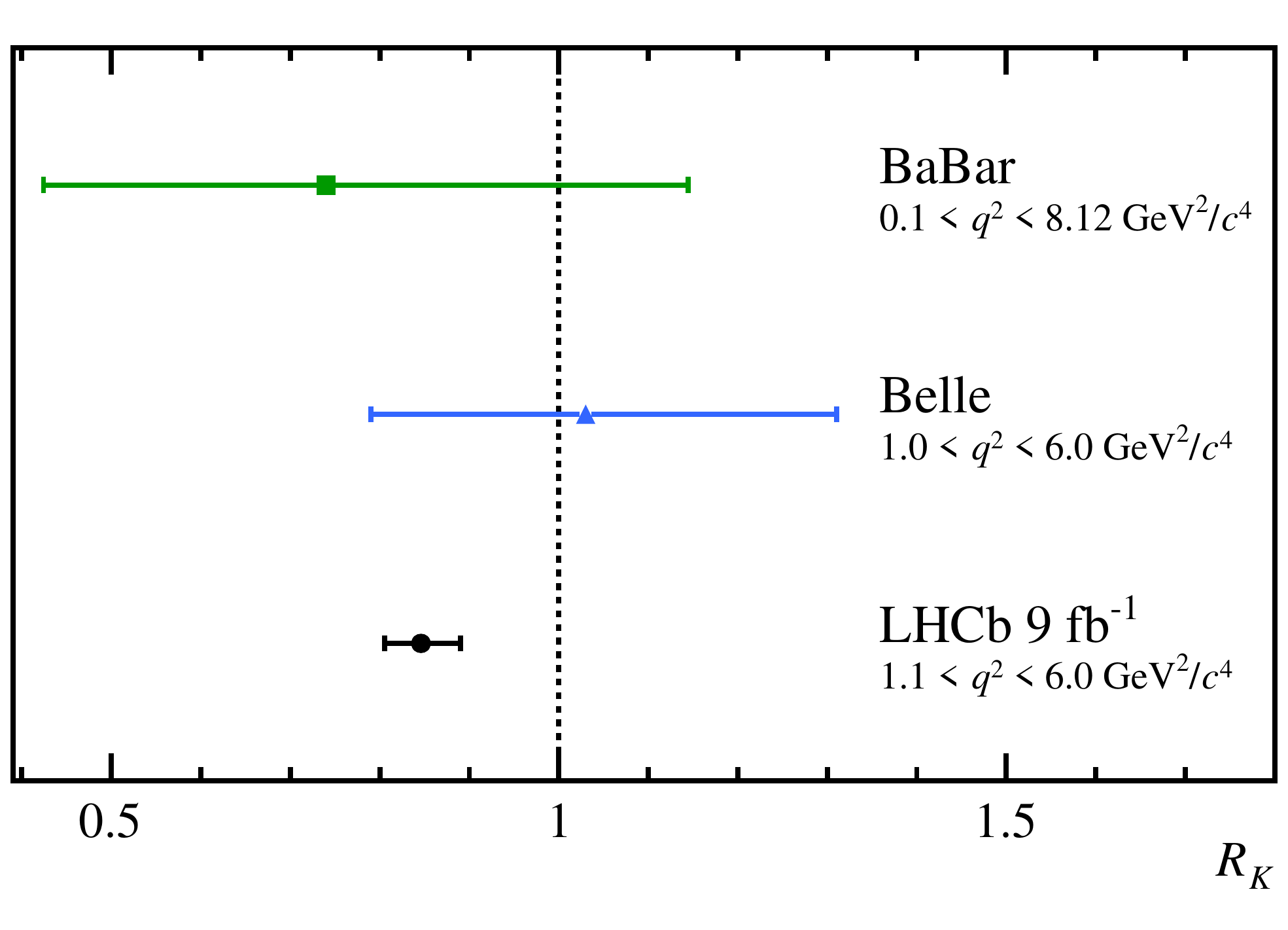}
\hspace{1em}
\includegraphics[width=0.41\linewidth]{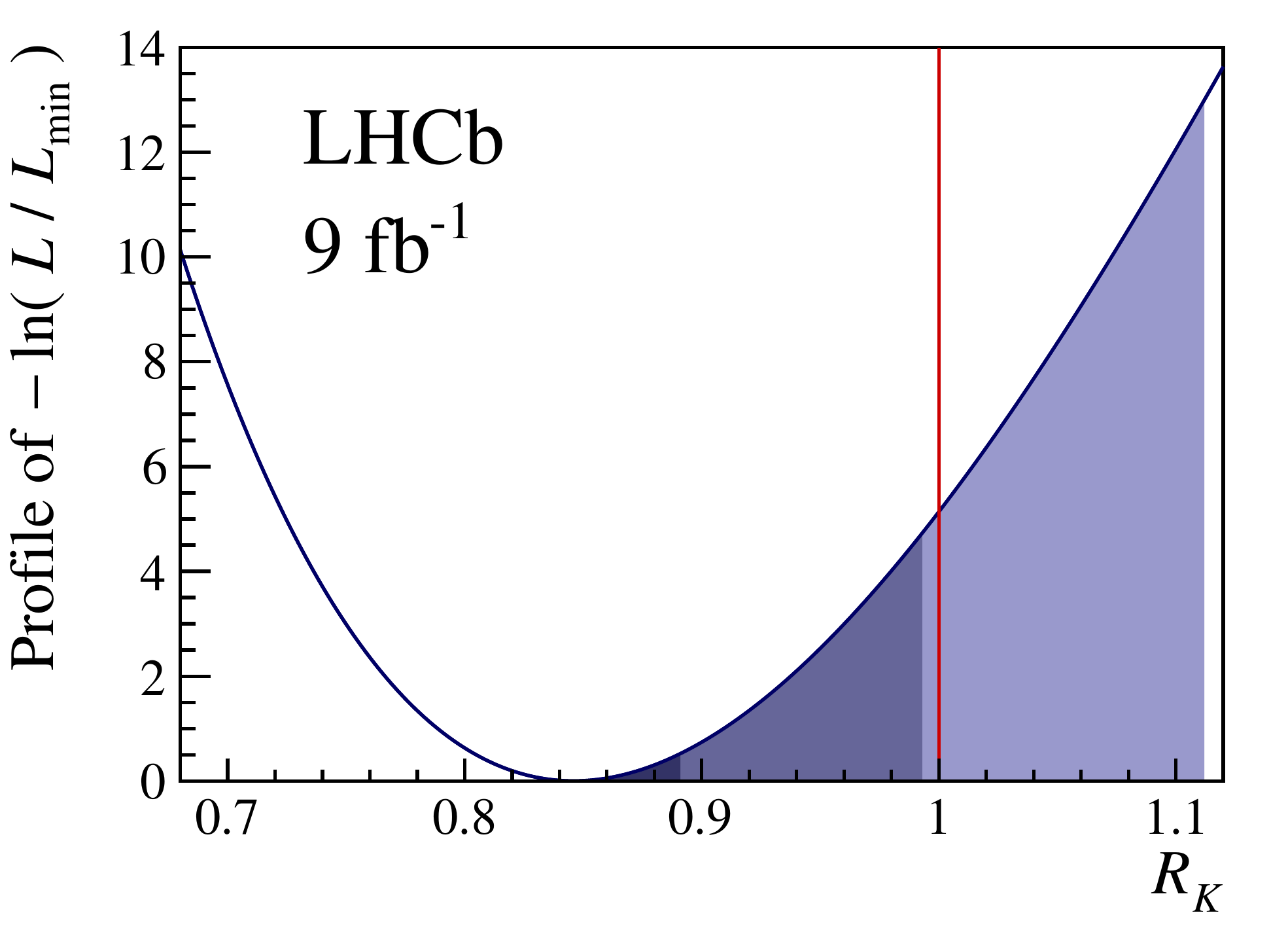}
\end{center}
\vspace*{-0.5cm}
\caption{
Left: The measurement of \RK presented in this article (black), compared to the SM prediction (dotted line) and the results from the BaBar~\protect\cite{RK_babar} (green) and Belle~\protect\cite{Abdesselam:2019lab} (blue) collaborations. Right: Negative log-likelihood from the fit to \BKll data, profiled as a function of \RK. Also shown are the SM prediction (red) and the values of \RK allowed at \SI{1}{\sigma}, \SI{3}{\sigma}, and \SI{5}{\sigma} levels (dark, medium, and light purple bands, respectively).
}\label{fig:RKresult}
\end{figure}

The final result is:

\begin{equation}
\label{eq:RKresult}
\RK=\RKresult,
\end{equation}

\noindent where the first uncertainty is statistical, and the second is systematic.
This is the most precise measurement of this quantity to date, as illustrated on the left-hand side of~\Cref{fig:RKresult}. 
The \mbox{$p$-value} under the SM hypothesis is evaluated based on the profiled log-likelihood, which is shown on the right-hand side of~\Cref{fig:RKresult}. This $p$-value is found to be $10^{-3}$, which corresponds to a significance of \RKsiggers. Therefore, the present \RK result provides evidence for the violation of LFU in \BKll decays.

The result for \RK is combined with the measurement of $\mathrm d\BR(\BKmumu)/\mathrm d\qsq$ from Ref.~\cite{Aaij:2014pli} to calculate the differential branching fraction of the \BKee decay, averaged over \linebreak \qsqRange. The result is:

\begin{equation}
\frac{\mathrm{d}\BR(\BKee)}{\mathrm{d}\qsq}= \left(28.6\,^{+\, 1.5}_{-\, 1.4}\,\text{(stat.)} \,\pm 1.3\,\text{(syst.)}\right)\times 10^{-9}\,\gev^{-2}\,.
% \right|_{\scriptsize \qsqRange}
\end{equation}

\section{Summary}\label{sec:conclusions}

This article covers the most precise measurement to date of lepton flavour universality in rare beauty-quark decays~\cite{Aaij:2021vac}. The observable \RK is measured using \SI{9}{\invfb} of proton-proton collision data recorded by the LHCb experiment. The result is found to be in tension with the SM prediction at the level of \RKsiggers. This measurement therefore provides evidence for the violation of LFU in \BKll decays. If confirmed by observations from independent sources, the breaking of LFU would require extensions to the SM that would accommodate lepton non-universal effects other than the Higgs interaction.

\section*{References}
\bibliography{main}

\end{document}